\documentclass[letterpaper]{llncs}
\usepackage[T1]{fontenc}
\usepackage[utf8]{inputenc}
\usepackage[table]{xcolor}
\usepackage{letltxmacro}
\usepackage{todonotes}
\usepackage{xspace}
\usepackage[hyphens]{url}
\usepackage{hyperref}
\usepackage{csquotes}
\usepackage{times}

\newif\ifabridged
\newif\ifnotabridged
\newif\ifanonymous
\newif\ifnotanonymous

\ifdefined\isabridged
\abridgedtrue
\fi

\ifabridged\notabridgedfalse
\else\notabridgedtrue
\fi

\ifdefined\isanonymous
\anonymoustrue
\fi

\ifanonymous\notanonymousfalse
\else\notanonymoustrue
\fi

\LetLtxMacro{\todonote}{\todo}
\renewcommand{\todo}[2][]
{\todonote[inline, caption={#2}, size=\footnotesize, #1]
{\renewcommand{\baselinestretch}{0.5}\selectfont#2\par}}

\title{On Making Emerging Trusted Execution Environments Accessible to Developers
\ifnotabridged{
\thanks{This is the author's version of the article to appear in 8th
Internation Conference of Trust \& Trustworthy Computing, TRUST 2015,
Heraklion, Crete, Greece, August 24-26, 2015. The final publication 
is available at \url{link.springer.com}}
}\fi}

\ifanonymous
\else
\author{Thomas Nyman\inst{1}\textsuperscript{,\textasteriskcentered}
  \and Brian McGillion\inst{1}\textsuperscript{,\textdagger}
  \and N. Asokan\inst{2}
}
\institute{Intel Collaborative Research Institute for Secure Computing (ICRI-SC)\\
  at Aalto University, Finland.
  \textsuperscript{\textasteriskcentered}\email{thomas.nyman@aalto.fi}
  \textsuperscript{\textdagger}\email{brian.mcgillion@aalto.fi}
  \and Aalto University and University of Helsinki, Finland.
  \email{asokan@acm.org}
}
\fi

\ifnotabridged{
\renewcommand\footnotemark{}
\fi

\begin{document}
\maketitle

\begin{abstract}
  New types of Trusted Execution Environment (TEE) architectures like
  TrustLite and Intel Software Guard Extensions (SGX) are
  emerging. They bring new features that can lead to innovative
  security and privacy solutions. But each new TEE environment comes
  with its own set of interfaces and programming paradigms, thus
  raising the barrier for entry for developers who want to make use of
  these TEEs.  In this paper, we motivate the need for realizing
  standard TEE interfaces on such emerging TEE architectures and show
  that this exercise is not straightforward. We report on our
  on-going work in mapping GlobalPlatform standard interfaces to
  TrustLite and SGX.
\end{abstract}

\section{Introduction}
\label{sec:intro}

For more than a decade the vast majority of smartphones and tablets have been equipped with hardware security functionality, usually referred to as \emph{Trusted Execution Environments } (TEEs)~\cite{EkbergKA14}. A TEE is an isolated and integrity-protected processing environment where sensitive computations, such as cryptographic operations, can be safely carried out. Until recently, application developers have not had the means to make use of TEEs to enhance the security and privacy of their applications. New standardization efforts, such as GlobalPlatform (GP)~\cite{web:GP_TEE}, and open source implementation initiatives, such as OP-TEE~\cite{web:linaro_op_tee} and Trusted Little Kernel~\cite{web:NVIDIA_TLK} have the potential for ushering in widespread use of TEEs by application developers.

Although the deployed base of mobile devices with TEEs is very large, numbering in hundreds of millions, they predominantly follow the same architectural pattern: a computing device containing a physical or logical TEE where small amounts of sensitive computation can be carried out in conjunction with a larger software components operating outside the TEE. The chief example of such a TEE architecture is ARM TrustZone~\cite{ARM09} which is widely deployed in smartphones.

Recently new types of TEE architectures have been proposed. They range from TEEs like TrustLite~\cite{Koeberl14} and SMART~\cite{EldefrawyTFP12} designed for tiny resource constrained devices to Intel Software Guard Extensions (SGX)~\cite{McKeen13} intended primarily for servers and desktops. They provide novel functionality but come with their own \emph{Application Program Interfaces} (APIs) and development environments. This constitutes a high barrier for entry for developers making it more difficult for them to benefit from such TEE functionality. A natural solution approach to this problem would be to support a set of standard TEE interfaces, such as those specified by GP, on these emerging TEE architectures thereby allowing developers familiar with the standard interfaces to readily make use of the new TEEs. However, subtle differences in architectural assumptions pose some challenges in mapping GP interfaces to these TEEs. In this work-in-progress paper, we examine these challenges.

Our contribution is two-fold. First, we briefly juxtapose the features of the emerging TEE architectures with the assumptions behind TEE standards (Section~\ref{sec:background}). We then show the subtleties and challenges in realizing the GlobalPlatform model on two such example TEE architectures: TrustLite and SGX (Section~\ref{sec:mapping}). We briefly describe our experience so far in resolving these challenges.

\section{Background}
\label{sec:background}

\begin{figure}
  \centering
  \includegraphics[width=.8\textwidth]{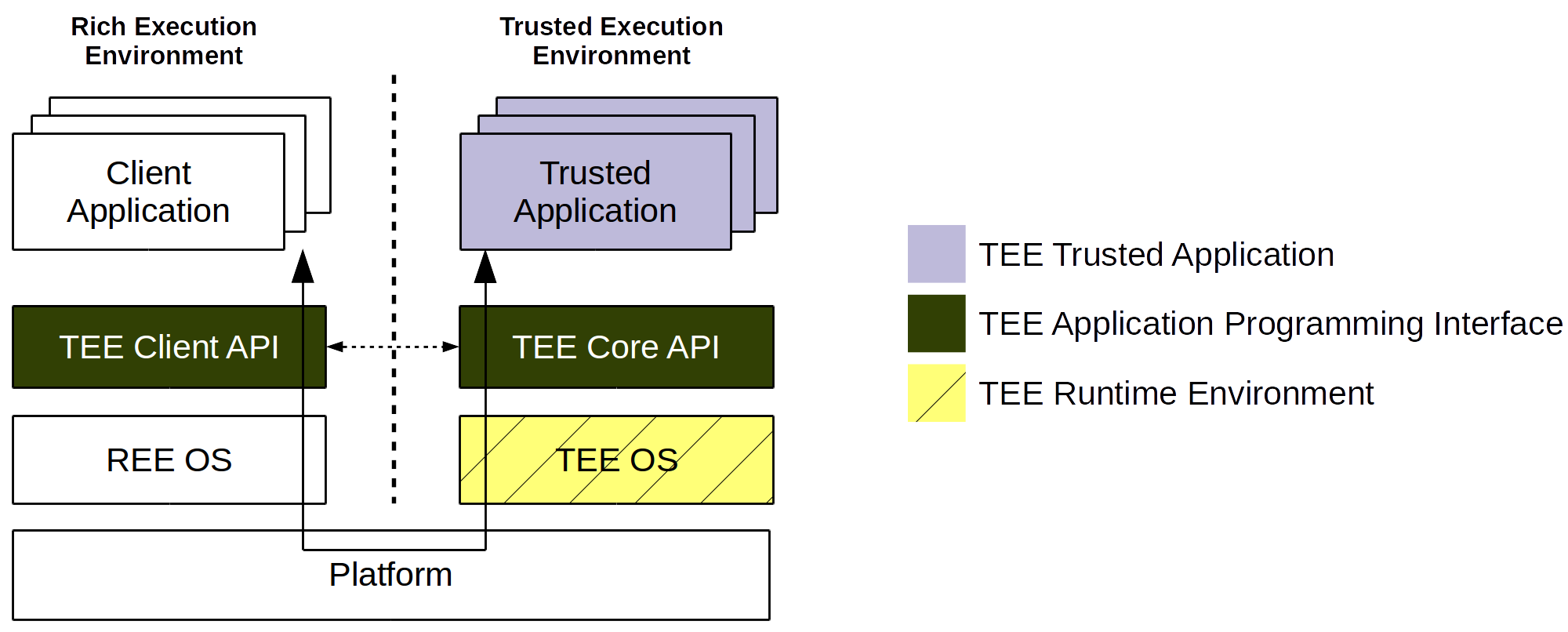}
  \caption{Abstract view of computing device equipped with a TEE~\cite{web:GP_TEE}}
  \label{fig:tee-arch}
\end{figure}

The isolation and integrity-protection of the processing environment in a TEE can be achieved in different ways. Contemporary TEE architectures commonly utilize either a dedicated security co-processor or CPU extensions that allow one physical processor to operate efficiently in two distinct isolated modes in a so called \emph{split-world} configuration. In either case, the device may be viewed as having two separate environments, each with its own set of features, as shown in Fig.~\ref{fig:tee-arch}. The \emph{Rich Execution Environment} (REE) refers to the operating environment that houses the conventional OS and applications. In contrast, the TEE typically only has a limited set of features that are only intended to address security critical functionality offloaded onto the TEE by \emph{Client Applications} (CAs) in the REE. The isolation of the TEE itself and discrete pieces of software inside the TEE, referred to as \emph{Trusted Applications} (TAs) is realized by means of hardware support. Services and APIs available to TAs, such as access to trusted storage, are provided by a TEE OS, part of the TEE \emph{Trusted Computing Base} (TCB).

\ifnotabridged{
A discrete co-processor is either be embedded on the main core physical \emph{System on Chip} (SoC), or a discrete hardware module outside the main core SoC. The advantages of a dedicated co-processor compared to split-world-based isolation is the capability to run simultaniously as the main core. A discrete co-processor can also enable a higher level of isolation, especially in the case of an external co-processor which does not necessarily share \emph{any} resources with the main core. On the other hand, co-processor architecture may exhibit increased overhead when transferring data to and from the co-processor. The co-processor is also generally less powerful compared to the main core.

In split-world-based isolated execution, such as ARM TrustZone~\cite{ARM09}, CPU state is banked between the two isolated modes; the \emph{normal world} and the \emph{secure world}. Transfer of execution between worlds occurs via a higher privilege \emph{monitor mode} that marshals the transition to and from the secure world. The secure monitor may be invoked through a \emph{Secore Monitor Call} (SMC) instruction. In addition, split-world-based isolation needs to be extended to both on-SoC and off-SoC memory and peripherals. In ARM TrustZone this protection is provided by TrustZone-aware SoC peripherals and a \emph{Advanced eXtensible Interface} (AXI) system bus which propagates an extra control signal known as the \emph{Non-Secure} (NS) bit system-wide via bus transactions. 

Physical memory partitioning between the normal and secure world is supported through the \emph{TrustZone Address Space Controller} (TZASC) and \emph{TrustZone Protection Controller} (TZPC). The TZASC can partition DRAM into several memory regions, each of which can be configured to be accessible from either the normal or secure world, or based on a more dynamic access control policy. On-SoC static memory like ROM or SRAM protection is provided by the \emph{TrustZone Memory Adapter} (TZMA) SoC peripheral. TZPC is mainly used to configure peripherals as secure or non-secure for access control on the low-bandwidth \emph{Advanced Peripheral Bus} (APB) bus attached to the AXI bus using a AXI-to-APB bridge. Unlike the AXI bus, the APB bus does not propagate the NS bit. The AXI-to-APB bridge mediates access to APB peripherals and will deny illegal access to protected peripherals. A TrustZone-aware \emph{Memory Management Unit} (MMU) provides both worlds with distinct translation tables. \emph{Translation Lookaside Buffer} (TLB) entries are tagged secure or non-secure, so that secure and non-secure data may coexists in the cache thus avoiding the need to flush the TLB when a world switch occurs. For \emph{Direct Memory Access} (DMA), the multi-channel \emph{Direct Memory Access Controller} (DMAC) which moves data around the physical memory system is made world sensitive and supports concurrent secure and non-secure channels. Any normal world DMA transfer to or from secure memory will be denied.

A split-world-based configuration trades off the need to offload data to and from the secure world with the cost associated with storing and restoring the world state when transfering execution between CPU modes. However, compared to a discrete co-processor, split-world-based isolation also results in a reduced cost per unit as the main core serves a dual purpose. Nevertheles, it should be noted that since the distinct worlds in fact share SoC resources, there may exist situations in which software in the normal world may interfere with the secure world indirectly (e.g. side-channels~\cite{Brumley15a}).
}\fi

\subsection{TrustLite and TyTAN}
\label{sec:trustlite}

TrustLite~\cite{Koeberl14} is a generic hardware security architecture intended for low-cost and exceedingly resource constrained embedded systems, such as automotive electronics, industrial control systems, medical implants or wearables. Such classes of devices have particularly strong economic incentives to minimize development and production costs, and hence typically lack hardware support for isolated execution or even support for paging and virtual memory. Nevertheless, in many cases such devices are employed in security-sensive and safety-critical applications, which not only require real-time guarantees, but can also benefit from certain security features, such as strong isolation and the possibility of attesting local state to a (remote) verifier.

Lacking a conventional \emph{Memory Management Unit} (MMU), memory access in these constrained environments is typically mediated by a \emph{Memory Protection Unit} (MPU), which can be programmed in supervisor mode by the device OS with memory access rules for the next task scheduled to run. In TrustLite the basis for strong isolation between tasks is an \emph{Execution Aware Memory Protection Unit} (EA-MPU), which not only enforces access control on all memory accesses, but does so considering the current program counter value when validating a particular access to memory. Hence, a platform equipped with an EA-MPU can enforce fine-grained access control based on the individual code regions executed independently of the OS. TrustLite also introduces a secure exception engine which maintains the memory isolation of secure tasks protected by the EA-MPU even in the case of hardware and software interrupts.

Later work by Brasser et. al. leverages the TrustLite architecture to realize TyTAN~\cite{Brasser15}, a security architecture for embedded systems that provides strong isolation of dynamically configurable tasks assisted by hardware features introduced in TrustLite, as well real-time scheduling guarantees. TyTAN utilizes the secure exception engine of TrustLite to provide a secure \emph{Inter Process Communication} (IPC) mechanism, where both sender and receiver can be authenticated using the digest of respective tasks measured upon loading. TyTAN does not provide any built-in access control for IPC. Instead, e.g., the task on the receiving end of an IPC call can make access control decisions based on the verified identity of the sender. TyTAN differs from the traditional TEE model in Fig.~\ref{fig:tee-arch} in that normal and secure tasks co-exist in a single environment\footnotemark, and rely on scheduling provided by an untrusted (real time) operating system. The OS is assumed to schedule tasks fairly without starvation. Normal tasks are isolated from other tasks but are fully controlled and accessible by the OS. In contrast, secure tasks are isolated by the EA-MPU from the rest of the system, including the OS. Whereas CAs in Fig.~\ref{fig:tee-arch} are thought to include the advanced domain logic of an application, and TAs only very limited security sensitive functionality, all tasks in TyTAN are relatively simple. The primary concern in TyTAN is the correct operation of secure tasks, even in the face of an adversary who has full control of the untrusted OS and normal tasks running on the platform.

\footnotetext{Instead, the class of devices TrustLite represents may potentially be used as part of a programmable secure co-processor.}

\subsection{Intel SGX}
\label{sec:sgx}

Intel Software Guard Extensions~\cite{McKeen13} is another hardware-based approach to realize an isolated environment for preserving the confidentiality and integrity of sensitive code and data. SGX consists of a set of new CPU instructions and memory access changes to the Intel CPU architecture, which allows parts of the application code and data residing in main memory to be encrypted using a key accessible only in the CPU core. The protected portions of the application's virtual address space together with the corresponding SGX control data structures are referred to as an \emph{enclave}. Enclave creation and initialization is handled via two dedicated instructions, \emph{Enclave CREATE} (\texttt{ECREATE)} and \emph{Enclave INIT} (\texttt{EINIT}). When an enclave is operated on, a dedicated hardware unit on the CPU package decrypts incoming, and encrypts outgoing traffic between the main memory and CPU package, so that sensitive code and data never leave the CPU package unencrypted. Transfer of execution into and out of an enclave is strictly controlled.

Entry to an enclave occurs via a dedicated \emph{Enclave ENTER} (\texttt{EENTER}) CPU instruction, which causes any cached page table translations overlapping with the protected address region of an enclave to be invalidated and transfers control to code inside the enclave. While the CPU is executing in \emph{enclave mode}, it has access to the protected pages belonging to the currently executing enclave, as well as any unprotected pages in the current processes's virtual address space. Accesses to protected pages belonging to other enclaves are prevented. Outside of enclave mode, access attempts by the CPU to enclave pages are treated as references to nonexistent memory, as are physical memory accesses by other agents, such as DMA access by capable disk drive controllers, graphics cards etc. Furthermore, \texttt{SYSENTER} and \texttt{SYSCALL} instructions are prohibited while in enclave mode, requiring the enclave to be exited before making system calls.

Exit from an enclave may occur either as a result of an \emph{Enclave EXIT} (\texttt{EEXIT}) instruction, or asynchronously, such as when exceptions or interrupts occur during enclave execution. In the latter case, an \emph{Asynchronous Enclave eXit} (\texttt{AEX}) event causes the processor state to be securely saved inside protected enclave memory and the CPU registers to be scrubbed in order to avoid leakage of sensitive data. Finally, the CPU leaves enclave mode. The enclave may subsequently be re-entered with an \emph{Enclave RESUME} (\texttt{ERESUME}) instruction, which restores the execution state of the enclave. In both cases, any cached page translations referring to the enclave's protected address region are cleared.

In contrast to conventional split-world-based isolated execution approaches, such as ARM TrustZone\ifabridged~\cite{ARM09}\fi, the SGX hardware architecture results in much simpler transitions to and from the secure CPU mode. There is no need to manually transfer data back and forth between a secure and normal world, as the isolated enclaves execute within the address space of the host process. In addition there is no need for a separate operating environment to provide further isolation between enclaves, as each enclave is already isolated from other processes in the system, including it's own host process and other enclaves.

\subsection{Standardization}
\label{sec:standardization}

To date there have been many proprietary APIs developed for TEEs. Almost every vendor of TEE technology has supplied their own implementations, with varying levels of complexity and functionality. Many of these solutions are aimed at the same market segments. They thus impact the consumers, namely \emph{Original Equipment Manufacturers} (OEMs) and operators in this case, in similar ways. This fragmentation has been partly to blame for the slow uptake in the use of TEE technology. OEMs find that they need to provide more engineering effort to support essentially the same functionality on different hardware platforms. With ever increasing demands for more services and growing awareness of end users' need for privacy protection, OS vendors such as Google are mandating that more of the security critical components of the OS are protected by hardware security mechanisms.

This fragmentation and the need for OEMs to pass \emph{Compliance Test Suites} (CTSs), designed to test that the platform protects the security critical components as mandated by the OS vendor, have led to the need to address this through some form of standardization. There have also been a number of efforts to address this by providing a standard operating system to run within the TEE, such as Nvidia's \emph{Trusted Little Kernel}~\cite{web:NVIDIA_TLK} (TLK) and Linaro's \emph{OP-TEE}~\cite{web:linaro_op_tee}. Though a common operating system is a good start to help adoption as it removes a lot of engineering effort, it does not go far enough to address the needs of application developers on platforms such as Android or iOS. To benefit these application developers there must be a consistent API by which TEE functionality can be accessed.

To this end \emph{GlobalPlatform} (GP)~\cite{web:GP_TEE}, a non-profit association focused on promoting the industry wide adoption of security standards, has been formulating and driving standards for TEE APIs. Within the scope of the TEEs that we describe in this paper, i.e. extensible TEEs that allow for the deployment of trusted third-party code, GP has defined two main standards. The \emph{TEE Client API}, running in the host operating system, and the \emph{TEE Core API}, running inside the TEE. This provides standardization of both the TAs (so they can be deployed in any GP compliant TEE irrespective of the HW vendor) and the CAs that run on the host operating system. \ifnotabridged{It is these CAs that provide the services to meet the corresponding CTS requirements of OS vendors.}\fi GP's architecture is built around the model defined in Fig.~\ref{fig:tee-arch} where a CA running in the REE creates one or more sessions to TAs in the TEE. This session establishes an effective \emph{Remote Procedure Call} (RPC) mechanism through which it is possible for the CA to invoke commands in the TA. Via a well defined API a TA is able to provide its services while also protecting itself from misuse. The first column in Table~\ref{tbl:sgx-ta-interface} summarizes the GP TEE Core API internal TA interface.

To date GP is the forerunner in TEE standardization -- numerous GP implementations are starting to emerge. Both TLK and OP-TEE provide some form of GP compliance, though the exact extent of the compliance is unknown as the GP compliance test suite is not readily available.

\subsection{Open-TEE}

\begin{figure*}[t]
  \centering
  \begin{minipage}[t]{0.45\linewidth}
    \centering
    \includegraphics[width=\textwidth]{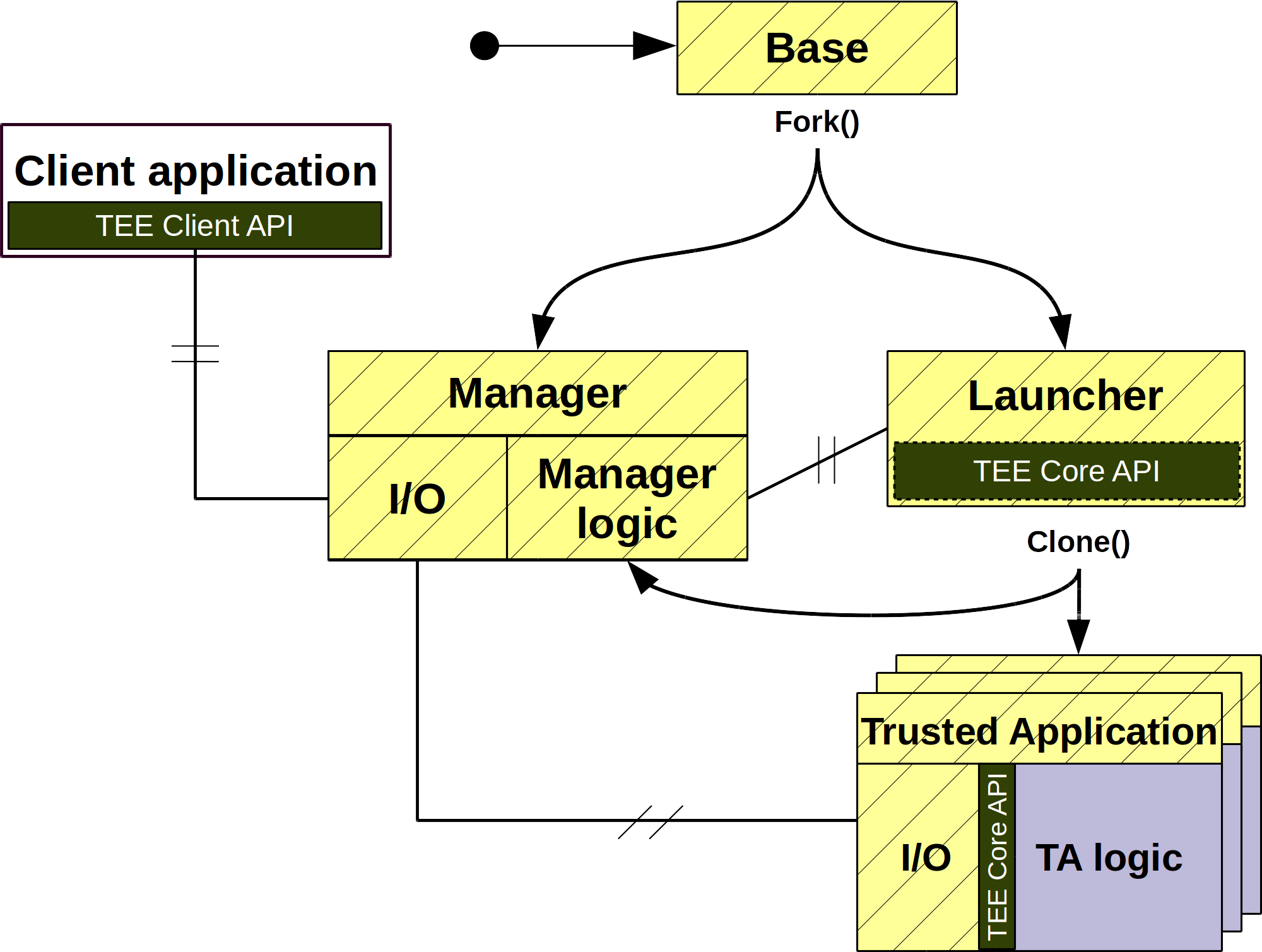}
    \caption{Open-TEE architecture~\cite{MDNA15}}
    \label{fig:opentee}
  \end{minipage}
  \hspace{12pt}
  \begin{minipage}[t]{0.45\linewidth}
    \centering
    \includegraphics[width=\textwidth]{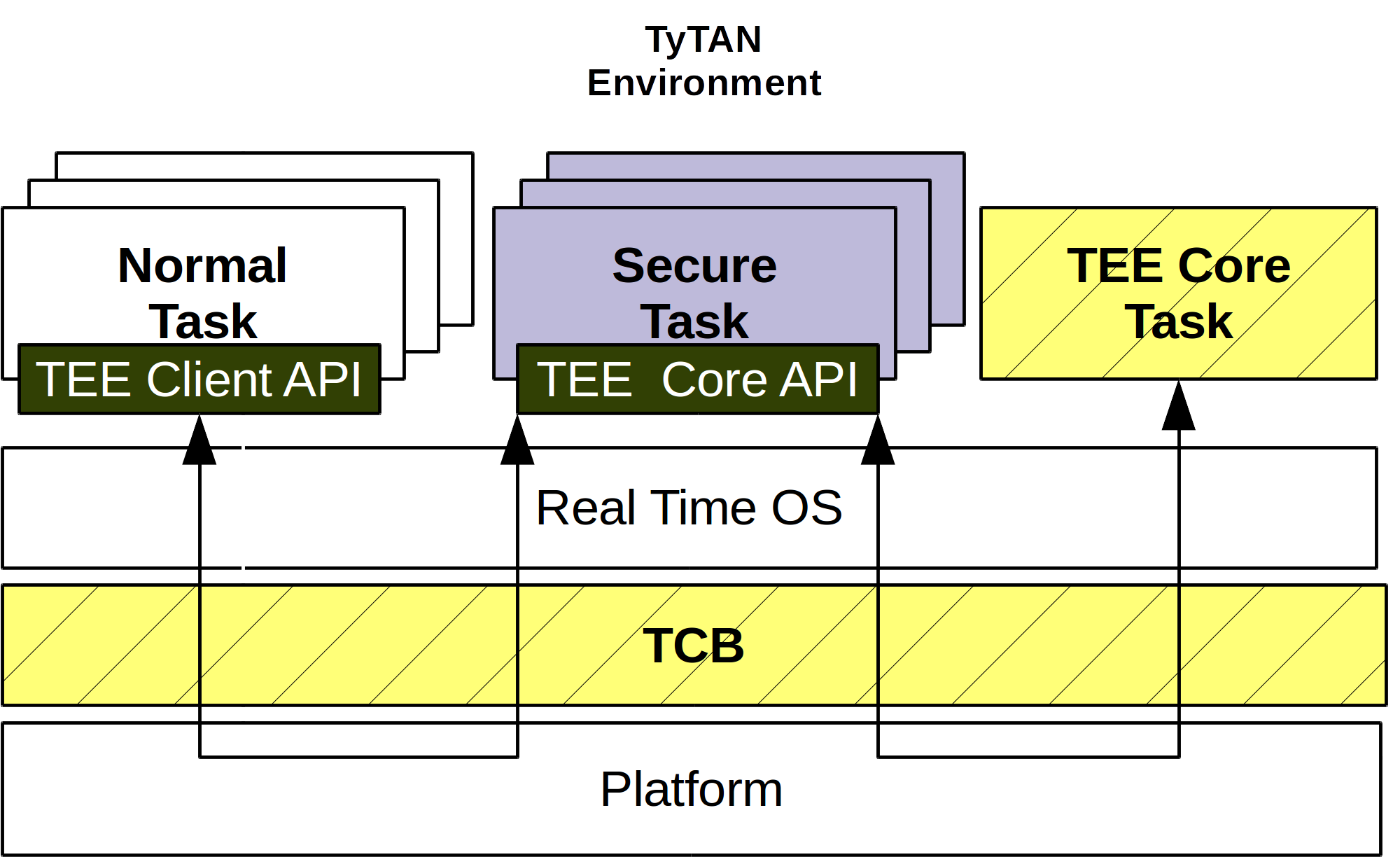}
    \caption{GP architecture on TyTAN}
    \label{fig:tytan-arch}
  \end{minipage}
\end{figure*}

\ifanonymous{McGillion et. al.~\cite{MDNA15} }\else{In our previous work~\cite{MDNA15}, while recognizing the efforts that have been undertaken to address the issue of TEE compliance, we }\fi highlight other issues that hinder widespread adoption of TEEs. Chief among them are the lack of access to debug-enabled versions of existing TEEs and the difficulty of developing applications for the TEE.
\ifnotabridged{

\begin{displayquote}
``Software development kits for TEE application development are often proprietary or expensive. Debugging low-level TEE applications either requires expensive hardware debugging tools, or leaves the developer with only primitive debugging techniques like “print tracing” (e.g., using printf statements in C to keep track of how values of variables change during program execution).''~\cite{MDNA15}
\end{displayquote}

}\fi
This insight led to the development of Open-TEE, a virtual TEE that complies with the GP standard and provides a fast and efficient prototyping platform for TAs / CAs. Fig.~\ref{fig:opentee} shows the Open-TEE architecture as a series of processes running on a development machine, thus allowing developers to leverage the tools, e.g. editors and debuggers, that they are familiar with. The \emph{Manager} process provides the TEE runtime with services usually expected from the TEE OS. Unlike a proper TEE OS, which would be self-contained and run on bare hardware, the implementation of Manager utilizes APIs provided by the host OS, in this case POSIX and a small number of APIs specific to Linux. The sole purpose of the \emph{Launcher} process is to pre-load the shared library implementing the TEE Core API and serve as a base process used to clone the actual TA processes, which are subsequently reparented onto Manager. Each TA process is divided into two separate threads; the \emph{I/O} thread and \emph{TA logic} thread. The I/O thread facilitates communication with Manager, whereas the TA logic thread executes the TA logic. \ifnotanonymous{Open-TEE is publicly available~\footnote{\url{http://open-tee.github.io/}} under the permissive Apache-V2 license.}\fi

\section{Mapping GlobalPlatform Interfaces to New TEE Architectures}
\label{sec:mapping}

Although not conventional TEE environments with their own OSs, TrustLite and SGX provide the same security assurances i.e. integrity and confidentiality of both code and data. They do not enforce the split world view that has become the de facto standard for extensible TEEs. However, there are still a number of use cases where applying the GP concepts and having access to a GP compliant implementation can be beneficial. First and foremost is the question of portability -- GP compliant applications should be readily deployable in any compliant TEE. Second is the notion of a services framework -- there are many cases when a developer would like to use existing services provided by a TEE, e.g. a keystore with key management and cryptographic routines, without the need to reimplement these.

\subsection{Realizing GP interfaces on TrustLite / TyTAN}

In Section~\ref{sec:trustlite} we noted how the isolation model in TyTAN differs from the typical GP model. Consequently, mapping the GP APIs to platforms such as TyTAN presents some challenges: 

\textbf{Placement of TEE Core API implementation} Typically, the TEE Core API and support for CA-to-TA communication would be provided as services by the TEE OS. In TrustLite and TyTAN the OS itself is excluded from the platform TCB; thus the TEE Core API implementation needs to be provided in some other manner. An obvious alternative would be to provide it purely in a library linked to each trusted task designated as a GP-style TA. This has a number of drawbacks. First, due the lack of paging memory management and caching, this leads to suboptimal memory utilization as identical functionality would be replicated in each TA. Second, the implementation of certain APIs, such as trusted storage access would be greatly simplified by platform support for centralized secure storage. Therefore TEE Core API functionality could either be provided directly by the TyTAN TCB, or in a less invasive approach, by a separate secure task, referred to as \emph{TEE Core task}, separately protected by the EA-MPU. The TEE Core API library linked to TAs can then invoke IPC calls to the TEE Core task, which resolves such calls appropriately. The overall architecture is shown in the Fig.~\ref{fig:tytan-arch}. In this case, the challenge stems from the fact that the strong isolation provided by TrustLite is not based on a completely separate environment, but on per-task isolation. Therefore, there is no separate trusted OS part of the TCB as is the case in established TEEs.

\textbf{Session access control} The TEE Client API is used for communication between normal and secure tasks. The TEE Core API also provides an Internal Client API, which allows one TA to act as a client to another TA. With regards to TyTAN tasks, this can be utilized for communication from one secure task to another. One challenge with this mapping is the need to enforce access restrictions on the TEE Core task so that only secure tasks are allowed to invoke TEE Core API functionality. In addition, the TEE Client API provides session-based access to TAs. As noted in Section~\ref{sec:trustlite}, TyTAN does not provide any IPC access control by itself, but leaves this up to individual secure tasks based on the hash digest of the sender as reported by the IPC primitives part of the TyTAN TCB. One approach would be to implement sessions, TEE Core task multiplexing and access control in each secure task individually. However, providing this functionality as part of a proxy in the TyTAN TCB has the same advantages as the centralized TEE Core placement which is an important consideration in the highly resource constrained environments that TyTAN targets. The challenge here stems from a large disparity in levels of IPC abstraction between TyTAN, and the GP specifications.

\ifnotabridged{
\textbf{Identifiers} One advantage of the TEE Client API compared to the secure IPC primitives in TyTAN is a more robust addressing scheme. TyTAN uses the hash digest of the intended receiver for secure IPC addressing. This creates a brittle binding between the sender and receiver, where a software update in the receiver necessitates update of the sender also to account for the new receiver digest. In scenarios involving communication between multiple tasks, the brittleness would cause a cascade of updates between tasks with dependencies between each other, making selective code update impossible. In contrast, TA addressing in the TEE Client API is based on \emph{Universally Unique IDentifiers} (UUIDs) assigned to TAs. However, as no centralized authority is required to administer UUID assignments, it would be desirable to bind a UUID to a set of valid task measurements, the verification of which may be delegated to the platform. The GP standards does not address this issue directly, but leaves TA verification up to individual implementations.
}\fi

\textbf{Other issues} Apart from the Internal Client API, the TEE Core API provides the TAs with internal programming interfaces for trusted storage, cryptographic operations, time API, and an arithmetic API intended as building blocks for developers to implement further asymmetric cryptographic algorithms. For performance reasons, platforms based on TrustLite may require additional cryptographic hardware accelerators to meet operational requirements, hence full coverage of the TEE Core API is not only impractical, but also unnecessary for many intended use cases. Due to its nature, it is likely that instantiations of TrustLite for different use cases will have varying degrees of hardware support for aforementioned features, and it is unlikely that a one-size-fits-all solution would be applicable for devices of this scale.

\subsection{Realizing GP with SGX through Open-TEE}

Open-TEE was conceived as a GP compliant tool for fast prototyping. However, throughout its design, choices were made that would allow it to work as a fully functional TEE environment when combined with the right hardware security mechanisms, like SGX enclaves. An enclave is a ring-3 construct. As such code in an enclave cannot make systems calls or other external interactions. Due to these restrictions, an enclave must synchronize with the non-enclave part of the application to perform external tasks on its behalf.

\begin{table}
\centering
\tt
\begin{tabular}{l l}
  \multicolumn{1}{c}{\textnormal{GP TEE Core API internal TA entry points}} & \multicolumn{1}{c}{\textnormal{SGX instructions and events}} \\
  \hline
  TA\_CreateEntryPoint & ECREATE, EINIT, EENTER, EEXIT \\
  \rowcolor{black!15} \cellcolor{black!15}TA\_DestroyEntryPoint & EENTER, EEXIT \\
  TA\_OpenSessionEntryPoint & EENTER, AEX, ERESUME, EEXIT \\
   \rowcolor{black!15} \cellcolor{black!15}TA\_CloseSessionEntry & EENTER, AEX, ERESUME, EEXIT \\
   TA\_InvokeCommandEntryPoint & EENTER, AEX, ERESUME, EEXIT \\
\hline
\end{tabular}
\normalfont
\caption{Mapping TEE Core API internal TA entry points to SGX instructions}
\label{tbl:sgx-ta-interface}
\end{table}

If we take a common usage scenario of a CA wishing to interact with a TA that performs some operation involving access to secure storage, we can see how Open-TEE's architecture (Fig~\ref{fig:opentee}) can be mapped to SGX~\cite{SGX_PROG}. The choice to split the TA process into two distinct threads, one handling I/O, and the other logic functionality facilitates this mapping. The I/O thread is responsible for all communication with Manager and CA. It is also responsible for all system interactions that may be required. The TA logic thread is where the TA code, which can make use of the TEE Core API, is executed. Manager provides secure storage functionality as a service to TA processes.

The CA initiates an open session call towards the TA. Manager noticing that the TA is not running requests Launcher to create a TA process. At this point the TA is a standard process and loads in accordance with the OS requirements. Once the TA is created we must initialize the TA logic code which conforms to the GP standard~\cite{web:GP_TEE} by invoking \texttt{TA\_CreateEntryPoint()}. As all of the GP conformant functionality is implemented within the enclave it is at this point that the enclave must be created. The TA hands over the enclave code along with any additional configuration data to the enclave creation service running in supervisor mode. Once the enclave has been created and initialized control is handed back to the TA application which enters the enclave to finalize the TA initialization and establish the session to the CA.

Now that the session is established between the CA and the TA, the CA can invoke TA commands using \texttt{TEEC\_InvokeCommand()}. Imagine that the CA wishes to provision some data to the secure storage. It invokes the corresponding TA command when the TA logic thread wants to access the secure storage. This requires a read or write operation to the storage media, which is a system call and is thus prohibited by SGX. In this case the Open-TEE implementation of the TEE Core secure storage API initiates an interrupt towards the I/O thread causing an \texttt{AEX} event which stores the enclave state and control is passed back to the I/O thread. The I/O thread then invokes the Manager to read/write encrypted data from/to persistent storage. When this action has been completed the TA can resume execution in the TA logic thread. When the invoked task is complete the enclave can be exited thereby returning control to the I/O thread which can then respond to the CA on the status of the invoked command. Other system services needed in the TEE Core API such as time functionality can be handled in a similar fashion. Table~\ref{tbl:sgx-ta-interface} summarizes the mapping of TEE Core API internal TA entry points to SGX instructions and events.

\section{Conclusion}

The GP TEE interfaces were designed to support split-world-based TEEs. We have shown that, even though potentially beneficial for easing the adoption of new, emerging TEE environments, mapping existing GP interfaces to them has a number of challenges. In ongoing work we are implementing the GP TEE Client API, and a subset of the TEE Core API on TyTAN. We are also exploring practical subsets of the TEE Core API applicable to different use cases. In future work we plan to investigate possible extensions to the TEE Core API relevant for small scale devices as well as realize Open-TEE on SGX.

\bibliographystyle{splncs03}
\bibliography{new-tee-challenges}

\begin{thebibliography}{10}
\providecommand{\url}[1]{\texttt{#1}}
\providecommand{\urlprefix}{URL }

\bibitem{ARM09}
{ARM Security Technology - Building a Secure System using TrustZone
  Technology}.
  \url{http://infocenter.arm.com/help/topic/com.arm.doc.prd29-genc-009492c/PRD29-GENC-009492C_trustzone_security_whitepaper.pdf}
  (2009)

\bibitem{Brasser15}
Brasser, F., et~al.: {TyTAN}: Tiny trust anchor for tiny devices. In: 52nd
  Design Automation Conference (DAC) 2015 (Jun 2015)

\bibitem{EkbergKA14}
Ekberg, J., Kostiainen, K., Asokan, N.: The untapped potential of trusted
  execution environments on mobile devices. {IEEE} Security {\&} Privacy
  12(4),  29--37 (2014), \url{http://dx.doi.org/10.1109/MSP.2014.38}

\bibitem{EldefrawyTFP12}
Eldefrawy, K., Tsudik, G., Francillon, A., Perito, D.: {SMART:} secure and
  minimal architecture for (establishing dynamic) root of trust. In: 19th
  Annual Network and Distributed System Security Symposium, {NDSS} 2012, San
  Diego, California, USA, February 5-8, 2012. The Internet Society (2012),
  \url{http://www.internetsociety.org/smart-secure-and-minimal-architecture-establishing-dynamic-root-trust}

\bibitem{web:GP_TEE}
GlobalPlatform: Device specifications for trusted execution environment.
  \url{http://www.globalplatform.org/specificationsdevice.asp}

\bibitem{SGX_PROG}
Intel: {Software Guard Extensions Programming Reference} (2013),
  \url{https://software.intel.com/sites/default/files/329298-001.pdf}

\bibitem{Koeberl14}
Koeberl, P., Schulz, S., Sadeghi, A.R., Varadharajan, V.: {TrustLite}: A
  security architecture for tiny embedded devices. In: Proceedings of the Ninth
  European Conference on Computer Systems. pp. 10:1--10:14. EuroSys '14, ACM,
  New York, NY, USA (2014), \url{http://doi.acm.org/10.1145/2592798.2592824}

\bibitem{web:linaro_op_tee}
Linaro: {OP-TEE}. \url{https://wiki.linaro.org/WorkingGroups/Security/OP-TEE}

\bibitem{MDNA15}
McGillion, B., Dettenborn, T., Nyman, T., Asokan, N.: {Open-TEE} -- an open
  virtual trusted execution environment. Tech. rep., Aalto University (2015),
  \url{http://arxiv.org/abs/1506.07367}

\bibitem{McKeen13}
McKeen, F., et~al.: Innovative instructions and software model for isolated
  execution. In: Proceedings of the 2nd International Workshop on Hardware and
  Architectural Support for Security and Privacy. pp. 10:1--10:1. HASP '13,
  ACM, New York, NY, USA (2013),
  \url{http://doi.acm.org/10.1145/2487726.2488368}

\bibitem{web:NVIDIA_TLK}
NVIDIA: {Trusted Little Kernel (TLK)}.
  \url{http://nv-tegra.nvidia.com/gitweb/?p=3rdparty/ote_partner/tlk.git;a=summary}

\end{thebibliography}

\end{document}